\documentclass[pra,onecolumn,12pt]{revtex4}

\usepackage{amssymb}
\usepackage{amsmath}
\usepackage{bm}
\usepackage{natbib}
\usepackage{graphics,graphicx}
\usepackage{epsfig}
\usepackage{latexsym}
\usepackage{epic,eepic,graphicx,amssymb,amsmath,indentfirst}
\usepackage{bm}
\usepackage{graphicx}
\usepackage[english]{babel}
\usepackage[latin1,utf8]{inputenc}
\usepackage{multirow}
\usepackage{bbold}
\usepackage{color}
\usepackage[colorlinks={true}]{hyperref}
\usepackage{bbold}
\hypersetup{citecolor={blue}, filecolor={blue}, linkcolor={blue}, urlcolor={blue}}
\newcommand{\bra}[1]{\langle #1 \vert}
\newcommand{\ket}[1]{\vert #1 \rangle}

\begin{document}

\title{\bf Optimal solutions to quantum annealing using two independent control functions}

\author{Marllos E. F. Fernandes}
\email{marllos@df.ufscar.br}
\author{Emanuel F. de Lima}
\email{emanuel@df.ufscar.br}
\author{Leonardo K. Castelano}
\email{lkcastelano@df.ufscar.br}
\affiliation{Departamento de F\'isica, Universidade Federal de S\~ao Carlos (UFSCar)\\ S\~ao Carlos, SP 13565-905, Brazil}
\date{\today}

\begin{abstract}
We investigate the quantum computing paradigm consisted of obtaining a target state that encodes the solution of a certain computational task by evolving the system with a combination of the problem-Hamiltonian and the driving-Hamiltonian.   
We analyze this paradigm in the light of Optimal Control Theory considering each Hamiltonian modulated by an independent control function. In the case of short evolution times and bounded controls, we analytically demonstrate that an optimal solution consists of both controls tuned at their upper bound for the whole evolution time. This optimal solution is appealing because of its simplicity and experimental feasibility. To numerically solve the control problem, we propose the use of a quantum optimal control technique adapted to limit the amplitude of the controls. As an application, we consider a teleportation protocol and compare the fidelity of the teleported state obtained for the two-control functions with the usual single-control function scheme and with the quantum approximate optimization algorithm (QAOA). We also investigate the energetic cost and the robustness against systematic errors in the teleportation protocol, considering different time evolution schemes. We show that the scheme with two-control functions yields a higher fidelity than the other schemes for the same evolution time.

\end{abstract}

\maketitle

\section{Introduction}

Quantum computing is currently one of the most exciting and prominent areas of Physics due to its possibility of a revolutionary change in present-day technologies \cite{Arute2019,Wright2019,Peruzzo2014}. However, rather than large-scale quantum computers, we have closer at our disposal the so-called Noisy Intermediate-Scale Quantum (NISQ) devices, which can already, in principle, surpass the power of classical computers \cite{Preskill2018quantumcomputingin}. In the current NISQ era, contrasting to the standard quantum computing based on a set of logic gates, new computational paradigms have emerged based on the evolution of suitably designed Hamiltonians. For instance, variational quantum algorithms can operate in NISQ platforms to handle a variety of problems, such as ground-state chemistry, machine learning and combinatorial optimization \cite{Kandala2017,farhi2014quantum,doi:10.1126/science.abb9811, 10.22331/qv-2020-03-17-32,Harrigan2021,PhysRevA.104.032401,PhysRevLett.127.120502}. 

Variational quantum algorithms (VQAs) are typically based on a properly parametrized time-dependent Hamiltonian, which is applied to a register of qubits \cite{Cerezo2021-jv,Medvidovic2021-gh}. At the final-time, the registers will contain the solution for the problem. VQA aims to solve a specific problem by varying a set of parameters of the Hamiltonian, which also parametrizes the evolution. In this context, quantum approximate optimization algorithms (QAOAs) has been developed for solving combinatorial optimization problems \cite{Willsch2020}. QAOA consists of applying a sequence of two unitary evolution operators originated from the so-called problem-Hamiltonian and driving-Hamiltonian in alternation, where the parameters correspond to the timings of these unitary operators.

Adiabatic quantum algorithms (AQAs), also in the class of quantum computing approaches that employ a quantum hardware expressed as time-dependent Hamiltonians, have preceded VQAs \cite{RevModPhys.90.015002,Barends2016}. AQAs exploit the fact that a quantum system remains in its ground state if the evolution is made sufficiently slow. In this approach, the system evolves adiabatically from the driving-Hamiltonian to the problem-Hamiltonian, whose ground state encodes the solution of the computational task. However, the time required to keep the adiabatic condition valid can be too long to be useful in practical situations. Several alternatives have been proposed to circumvent this problem, such as the local adiabatic evolution or the use of counter-diabatic drivings \cite{PhysRevApplied.15.024038,PhysRevResearch.3.013227}. Related to AQA, the method of quantum annealing (QA) emerged as a quantum version of the classical optimization technique of simulated annealing \cite{Crosson2021,PhysRevE.58.5355,doi:10.1126/science.1068774,Boixo2014}. As in AQA, the connection of the initial driving-Hamiltonian to the problem-Hamiltonian in QA is carried out through a smooth continuous function of time, but allowing for faster, nonadiabatic dynamics.

Recently, several works have explored connections of optimal control theory with VQA and also with QA \cite{PhysRevResearch.3.023092,PhysRevA.100.022327,PhysRevA.91.043401,Isermann2021}. In Ref.\cite{PhysRevX.7.021027}, it has been argued that for a fixed time, square pulse controls (bang-bang type) are optimal, giving support to the QAOA methodology. However, from the standard form of QA involving a single control function, Ref.~\cite{PhysRevLett.126.070505} has shown that a more general form of the optimal solution is of "bang-annealing-bang" type, meaning that an hybrid time-dependent control starting at the minimum allowed value and ending at the maximum possible value, with an smooth annealing segment in between is usually optimal. In Ref.~\cite{PRXQuantum.2.010101} a comprehensive perspective of the possible connections between VQA with quantum control has been presented. Specifically, the authors indicate that the quantum optimal theory background can extract a rich variational structures of VQA and provide a better understanding of variational experiments.

 In the present work, we formulate the quantum computing problem of reaching the target state with both the driving- and the problem- Hamiltonians modulated by independent control functions. We analyze this problem in the framework of Optimal Control Theory. For bounded controls, there are two regimes depending on the total evolution time: for sufficiently long times, optimal solutions exist such that the target ground state can be exactly obtained; whereas, for short times, optimal solutions cannot fully reach the target state. We demonstrate that when the final time is so short that the target state cannot be fully prepared, the optimal solutions are those with both controls set to their maximum values during the whole evolution time. Moreover, we show that the use of two-control functions can generate higher yields than standard QA and QAOA schemes for the same evolution time.
 
 Numerically, the optimal controls are sought by the two-point boundary-value quantum control paradigma (TBQCP)~\cite{Taksan}, a monotonic quantum control method, which we have adapted to limit the amplitudes of both controls. The TBQCP technique has been employed to study protocols related to standard-gate quantum computing with great efficiency, {\it e. g.}, to compute  optimal controls capable of implementing the universal set of quantum gates in double quantum dots~\cite{QDS} or the permutation algorithm in hybrid qubits~\cite{permutation}.
 
 We apply the two-control methodology to implement a quantum teleportation protocol. Teleportation allows for the transmission of unknown states between two agents separated by a long distance \cite{PhysRevLett.103.120504,PhysRevA.93.012311}. Also, teleportation is an interesting example to investigate because it can be used to implement universal quantum computing.~\cite{PhysRevLett.103.120504}. Our results show that the two-control functions perform the quantum teleportation in the shortest time when their amplitude is not limited, but at a high energy cost. On the other hand, the two-control functions with limited amplitude achieves the desired goal of reaching the target state in a shorter time when compared to the single control-function QA with limited amplitude and to the QAOA approaches with a similar energy cost found in the other time evolution schemes. Moreover, we investigate the robustness against systematic errors considering different schemes to perform the teleportation and we verify that the two-control functions is more robust than the other tested schemes.
 

\section{Setup of the control problem}

Consider that we are given two time-independent Hamiltonians: $H_0$ (driving) and $H_1$ (problem), with $\ket{\phi_0}$ being the ground state of $H_0$ and $\ket{\chi_0}$ the ground state of $H_1$. Suppose that the system is initially prepared in the ground state of $H_0$, $\ket{\psi(t=0)}=\ket{\phi_0}$, and evolves according to the total time-dependent Hamiltonian,
\begin{equation}
    H(t)= \varepsilon_0(t)H_0+\varepsilon_1(t)H_1\label{Ht},
\end{equation}
where $\varepsilon_0(t)$ and  $\varepsilon_1(t)$ are two independent control functions. 

We seek to find the controls $\varepsilon_0^*(t)$ and $\varepsilon_1^*(t)$ that maximize the expectation value of a given observable $O$ at the final time $t=T$, expressed as a functional of the controls, 

\begin{equation}\label{ofunctional}
 J[\varepsilon_0,\varepsilon_1]=\langle O(T) \rangle\equiv\bra{\psi(T)}O\ket{\psi(T)}, 
\end{equation}
limiting the amplitudes of the controls such that $0\leq\varepsilon_k\leq1$, $k=0,1$.

We point out that in the standard versions of adiabatic/annealing computing, the controls $\varepsilon_k$ are not independent, being related by $\varepsilon_1(t)=1-\varepsilon_0(t)$, while the observable is usually the projection onto the ground state of the problem- Hamiltonian, $O=\ket{\chi_0}\bra{\chi_0}$. By the adiabatic theorem, it is known that an optimal solution to the control problem exists for sufficiently long final times $T$.  

It is convenient to consider the problem in terms of the maximization of the control Hamiltonian $\mathcal{H}$ (not to be confused with the system Hamiltonian $H(t)$) given by \cite{kirk},
\begin{equation}\label{cHdef}
    \mathcal{H}(\psi,\lambda,\varepsilon_0,\varepsilon_1,t)=\bra{\lambda(t)}-{\rm i} H(t)\ket{\psi(t)}+\mathrm{c.c.},
\end{equation}
where $\ket{\lambda(t)}$ is a Lagrange multiplier.

Substituting Eq.(\ref{Ht}) into (\ref{cHdef}) we obtain,
\begin{equation}\label{chamil}
    \mathcal{H}(\psi,\lambda,\varepsilon_0,\varepsilon_1,t)=\varepsilon_0(t)\Phi_0(t)+\varepsilon_1(t)\Phi_1(t),
\end{equation}
where
\begin{equation}\label{gradients}
    \Phi_k(t)\equiv2\operatorname{Im}{\left\{\bra{\lambda(t)}H_k\ket{\psi(t)}\right\}},\;\;k=0,1.
\end{equation}

According to the Pontryagin's maximum principle, necessary conditions for an optimal solution are obtained from the control Hamiltonian as
\begin{subequations}\label{necond}
\begin{align}
   & \ket{\dot{\psi}^*(t)}=\frac{\partial \mathcal{H}^*}{\partial \bra{\lambda(t)}}  =-iH^*(t)\ket{\psi^*(t)},\; \ket{\psi^*(0)}=\ket{\phi_0}, \label{subeq1} \\
    & \ket{\dot{\lambda}^*(t)}=-\frac{\partial \mathcal{H}^*}{\partial \bra{\psi(t)}}  =-iH^* (t)\ket{\lambda^*(t)},\;  \ket{\lambda^*(T)}=O\ket{\psi^*(T)} \label{subeq2}
\\
    & \mathcal{H}(\psi^*,\lambda^*,\varepsilon_0^*,\varepsilon_1^*,t)  \geq \mathcal{H}(\psi^*,\lambda^*,\varepsilon_0,\varepsilon_1,t), \forall \textrm{ admissible } \varepsilon_0,\, \varepsilon_1, \label{subeq3}
 \end{align}
\end{subequations}
where the symbol $*$ refers to the dynamical quantities calculated at an optimal solution. These conditions are, in fact, necessary, although not generally sufficient, for any extremal control.

Considering admissible variations $\delta\varepsilon_k$ around the optimal solution, $\varepsilon_k(t)=\varepsilon_k^*(t)+\delta\varepsilon_k(t)$, Eq.~(\ref{subeq3}) yields,
\begin{equation}\label{eqvarcontrol}
    \frac{\partial\mathcal{H}(\varepsilon_k^*)}{\partial\varepsilon_k}\delta\varepsilon_k(t)=\Phi_k^*(t)\delta\varepsilon_k(t)\leq0,\;\;k=0,1.
\end{equation}
Thus, for each value of $k$ and for a given $t$, three situations can occur: (i) $\Phi_k^*(t)=0$, (ii) $\Phi_k^*(t)>0$ and $ \delta\varepsilon_k(t)<0$ , or (iii) $\Phi_k^*(t)<0$ and $ \delta\varepsilon_k(t)>0$. Note that $ \delta\varepsilon_k(t)<0$ for all admissible variations implies $\varepsilon_k^*(t)=1$, because the variations about the maximum value must be negative, while $ \delta\varepsilon_k(t)>0$ implies $\varepsilon_k^*(t)=0$, because the variations about the minimum must be positive. Note that when no bounds are imposed to the controls, both gradients must vanish at an optimum solution, $\Phi_0^*(t)=\Phi_1^*(t)=0$, because the signs of $ \delta\varepsilon_k(t)$ are arbitrary.

 In addition, since, $ \varepsilon_0(t)\dot{\Phi}_0=-\varepsilon_1(t)\dot{\Phi}_1$,  the time derivative of the control Hamiltonian is given by,
\begin{equation}
    \frac{d\mathcal{H}(t)}{dt}=\dot{\varepsilon}_0(t)\Phi_0+\dot{\varepsilon}_1(t)\Phi_1.
\end{equation}
For optimal solutions, either the control is constant, $\dot{\varepsilon}_k^*(t)=0$, or the gradient vanishes, $\Phi_k^*(t)=0$, which implies that the control Hamiltonian at an optimum $\mathcal{H}^*$ is a constant of motion. Also note that when the amplitudes of optimal controls are within the boundaries $0<\varepsilon_k^*(t)<1$, for all $t\in[0,T]$, the situation is similar to the unbound control problem: necessarily the gradients are null and the control Hamiltonian vanishes identically $\mathcal{H}^*=0$.    

The necessary conditions for an optimal solution evidence that when the final time $T$ is long enough, there are optimal solutions such that the control Hamiltonian $\mathcal{H}^*$ is null and the maximum possible value of the functional (\ref{ofunctional}), $J^*_{\rm max}$, can be reached \cite{PhysRevLett.126.070505}. On the other hand, below a certain critical value $T_c$ of the final time, $J_{\rm max}^*$ cannot be reached. In this case, $T<T_c$, the control Hamiltonian at the optimal solution is a positive constant. To show this, consider that the final time $T$ of the control problem is free and set a new cost functional $J'$ given by,

\begin{equation}
    J'[\varepsilon_0,\varepsilon_1,T]=J[\varepsilon_0,\varepsilon_1]-\alpha T,
\end{equation}
where $\alpha\geq0$ a given constant. The functional to be maximized, $J'$, has a penalty term to account for increasing the final time. We can associate the same control Hamiltonian (\ref{chamil}) to this free final-time cost functional, along with  (\ref{necond}), and the additional necessary condition,

\begin{equation}
    \mathcal{H}^*(T)=\alpha.
\end{equation}
Hence, the value of the control Hamiltonian at an optimal solution depends exclusively on the final time $T$. For $\alpha=0$, corresponding to a free final time with no penalty term, the control Hamiltonian calculated at an optimal solution vanishes for all times and $J^*$ assumes its maximum possible value $J^*=J_{\rm max}^*$. In this case, there will be a sufficiently long final time such that we can find optimal solutions such that the controls are within the boundaries. However, for $\alpha\neq0$, the control Hamiltonian at an optimal solution is a positive constant and $J_{\rm max}^*$ is not reached.

Returning to problem with a fixed final time, consider the case where $T<T_c$, meaning that the gradients $\Phi_k(t)$ do not vanish simultaneously and $\mathcal{H}^*=\textrm{constant}>0$. More specifically, set $O= \ket{\chi_0}\bra{\chi_0}$ which leads to $J^*<1=J^*_{\rm max}$. Furthermore, from Eqs.~(\ref{subeq1}-\ref{subeq2}) and (\ref{gradients}) we obtain

\begin{equation}
  \Phi_1^*(T)=2\operatorname{Im}{\left\{\bra{\lambda^*(T)}H_1\ket{\psi^*(T)}\right\}}\propto \operatorname{Im}{\left\{\left|\bra{\psi^*(T)}\chi_0\rangle\right|^2\right\}}=0
\end{equation}
and
\begin{equation}
  \Phi_0^*(0)=2\operatorname{Im}{\left\{\bra{\lambda^*(0)}H_0\ket{\psi^*(0)}\right\}}\propto \operatorname{Im}{\left\{\bra{\lambda^*(0)}\phi_0\rangle\right\}}=\operatorname{Im}{\left\{\left|\bra{\psi^*(T)}\chi_0\rangle\right|^2\right\}}=0,
\end{equation}
where we have used the fact that $\frac{d}{dt}\bra{\lambda^*(t)}\psi^*(t)\rangle=\bra{\dot{\lambda}^*(t)}\psi^*(t)\rangle+\bra{\lambda^*(t)}\dot{\psi}^*(t)\rangle=0$.

Thus,

\begin{equation}
    \mathcal{H}^*=\varepsilon_0^*(T)\Phi_0^*(T)=\varepsilon_1^*(0)\Phi_1^*(0).
\end{equation}

Since $\mathcal{H}^*$ is positive, we conclude that $\Phi_0^*(T)>0$ and $\Phi_1^*(0)>0$. Consequently, to maximize $\mathcal{H}^*$, we must have $\varepsilon_0^*(T)=\varepsilon_1^*(0)=1$, hence

\begin{equation}\label{chamilvalue}
   \mathcal{H}^*=\Phi_0^*(T)=\Phi_1^*(0).
\end{equation}

Moreover, the time derivative of the sum of the gradients can be written as,
\begin{equation}\label{timephi}
    \frac{d}{dt}\left(\Phi_0(t)+\Phi_1(t) \right)=\left(\varepsilon_0(t)-\varepsilon_1(t)\right)\xi(t) ,
\end{equation}
where $\xi(t)=2\operatorname{Re}\left\{\bra{\lambda(t)}\left[H_1,H_0\right]\ket{\psi(t)}\right\}$.

Therefore, a solution to the bounded control problem for $T<T_c$ is given by the condition where both controls are set to their (equal) maximum values $\varepsilon_0^*(t)=\varepsilon_1^*(t)=1$ $\forall t \in [0,T]$ is optimal since in this case Eqs. (\ref{chamilvalue}) and (\ref{timephi}) imply that $\mathcal{H}^*=\Phi_0^*(t)+\Phi_1^*(t)=\Phi_0^*(T)$.

Finally, it is worth comparing the yield for the same final time $T$ obtained with two-control functions with that of the standard QA scheme, which employs a single control function such that $\varepsilon_1(t)=1-\varepsilon_0(t)$. In particular, we note that an optimal solution for the two-control case such that $\varepsilon_0^*(t)=\varepsilon_1^*(t)=1$ cannot be an optimal solution of the single-control case. Thus, from condition (\ref{subeq3}), an optimal solution for the single-control case is, in general, a sub-optimal solution for the two-control case, i.e., $J_{\rm sc}^*\leq J^*$, where $J_{\rm sc}$ is the cost functional for the single-control case for the same final time $T$.

\subsection*{TBQCP optimization method}

In this section, we address the numerical optimization technique that we have adapted to impose bounds to the two-control functions. The TBQCP is an iterative monotonic method able to find optimal controls that, given an initial state $\ket{\psi(t=0)}\equiv\ket{\psi_0}$, maximize the expectation value of a physical observable $O$ at the final time $T$. Starting with some trial controls, the physical observable is evolved backwards (from the final time $t=T$ to the initial time $t=0$) through the following equation 
\begin{equation}\label{oper}
i\text{\ensuremath{\hbar}} \frac{\partial O^{(n)}(t)}{\partial t}=\left[O^{(n)}(t),\, H^{(n)}(t)\right],\; O(T) \rightarrow O(0),
\end{equation}
where $H^{(n)}(t)=\sum_{k=0}^1\varepsilon^{(n)}_k(t)H_k$ and $\varepsilon^{(n)}_k(t)$ are the controls in the nth iteration. The initial state $| \psi(0) \rangle$ is evolved forward with the time-dependent Schr\"odinger equation,
\begin{equation}\label{SEqu}
 i\text{\ensuremath{\hbar}}\frac{\partial|{\psi^{(n+1)}(t)}\rangle}{\partial t}=H^{(n+1)}(t)|\psi^{(n+1)}(t)\rangle,
 \end{equation} 
where  $H^{(n+1)}(t)=\sum_{k=0}^1\varepsilon^{(n+1)}_k(t)H_k$ and $\varepsilon_k^{(n+1)}(t)$ is the (n+1)st iteration control for $k=0,1$, which is calculated through the following expression
\begin{equation}
\varepsilon_k^{(n+1)}(t)=\varepsilon_k^{(n)}(t)+ \eta f^{(n+1)}_{k}(t),\;k=0,1,\label{fieldn}
\end{equation}
where $\eta$ is a positive constant and the control correction is given by
\begin{equation}
f^{(n+1)}_{k}(t) =\frac{2}{\hbar}\textrm{Im}\left\{ \langle \psi^{(n+1)}(t)| O^{(n)}(t) H_k| \psi^{(n+1)}(t)\rangle \right\},\;k=0,1.\label{fmu}
\end{equation}
It has been shown that, as an optimal solution is approaching, the control correction $f^{(n+1)}_{k}(t)$ is equivalent to the gradient $\Phi_k(t)$ (Eq.~(\ref{gradients})~\cite{Taksan}, which establishes a close connection between the TBQCP and the variational version of the optimal control theory.
Equations~(\ref{oper}-\ref{fieldn}) are solved in a self-consistent way, starting with  trial controls $\varepsilon_k^{(0)}(t)$ ($k=0,1$) and monotonically increasing the value of the desired physical observable $\langle O(T)\rangle=\langle\psi(T)| O(T)|\psi(T)\rangle$, see more details in Ref.~\cite{Taksan}. In addition, at each iteration step, the amplitudes of the controls are bounded to the interval $[0,1]$ by enforcing the value of the corresponding limit whenever the control transcend the bounds of the interval $[0,1]$ in Eq.(\ref{fieldn}). In this case, if the control function $\varepsilon_k^{(0)}(t)$ ($k=0,1$) exceed one of the bounds 0 or 1 in a certain time interval, its value is set equal to the crossed bound in this same time interval during the whole self-consistent calculation.

\section{Quantum approximate optimization algorithm}

To compare with the optimized solutions of the control problem we also consider the QAOA approach. In the spirit of VQAs, the QAOA ansatz consists of alternately switch-on each Hamiltonian $H_k$ ($k=0,1$) in a certain time interval, while the other one is switch-off. Thus, the time-evolution becomes
\begin{equation}
    |\psi(t)\rangle=U(H_0,\beta_p)U(H_1,\gamma_p)\ldots U(H_0,\beta_1)U(H_1,\gamma_1)|\psi(0)\rangle,\label{QAOA}
\end{equation}
where $U(H_0,\beta_j)=\exp(-{\rm i}H_0\beta_j)$ and $U(H_1,\gamma_j)=\exp(-{\rm i}H_1\gamma_j)$ are evolution operators. $\beta_j$ and $\gamma_j$ are real positive variational parameters that should be adjusted to maximize the expectation value of an observable $O$ at the final time $t=T$, $\langle O(T) \rangle$. The variational parameters are constrained such that $T=\sum_{j=1}^p(\gamma_j+\beta_j)$, where $p$ specifies the number of jumps from one Hamiltonian to another. By construction, QAOA provides a solution to the control problem such that the controls satisfy $\varepsilon_1(t)=1-\varepsilon_0(t)$ and $\varepsilon_0(t)$ is of bang-bang type, meaning that the control is turned on and off (possibly many time) during the evolution. We employ the particle swarm optimization for fixed values of $p$ to find the set of parameters $\{\gamma_j,\beta_j\}$~\cite{PSO,Sousa-FerreiraMAR}. In the same spirit of Ref.~\cite{PhysRevLett.126.070505}, QAOA will be used as a benchmark to check the efficiency of the other different schemes employed in this work.

\section{Teleportation protocol}

As an application, we consider the adiabatic gate teleportation protocol introduced in Ref.~\cite{PhysRevLett.103.120504} adapted to the two independent control functions. In this protocol, there are three qubits under the action of the Hamiltonian (\ref{Ht}), where the driving-Hamiltonian is
\begin{equation}
  H_0=-\hbar\omega_0\left(\sigma^2_x\sigma^3_x+\sigma^2_z\sigma^3_z\right),  
\end{equation}
whose ground state is twofold degenerated $|\phi^{(1)}_0\rangle=|0\rangle\otimes|\Phi\rangle$ and $|\phi^{(2)}_0\rangle=|1\rangle\otimes|\Phi\rangle$, where $\ket{\Phi}$ is a Bell state, $|\Phi\rangle=\left(|00\rangle+|11\rangle\right)/\sqrt{2}$, and $\sigma^j_m$ is the Pauli spin matrix in the $m$-direction acting on the $j$th-qubit. 

The problem-Hamiltonian is
\begin{equation}
 H_1=-\hbar\omega_0\left(\sigma^1_x\sigma^2_x+\sigma^1_z\sigma^2_z\right),
\end{equation}
whose ground state is also a twofold degenerated state given by $|\chi^{(1)}_0\rangle=|\Phi\rangle\otimes|0\rangle$ and $|\chi^{(2)}_0\rangle=|\Phi\rangle\otimes|1\rangle$. Thus, this protocol aims at teleporting the information initially encoded into the first qubit to the third qubit at the final time of evolution.    
For the application of the TBQCP method and the QAOA, we consider the target observable as the projection on the ground state of the problem-Hamiltonian, $O=\ket{\chi_0}\bra{\chi_0}$, which can be chosen as any linear combination of states $\ket{\chi_0^{1}}$ and $\ket{\chi_0^{2}}$ without loss of generality~\cite{PhysRevLett.103.120504}.
The ultimate goal is then to maximize the fidelity defined by
\begin{equation}
    F(T)=\langle O(T) \rangle=|\langle\psi(T)|\chi_{0}\rangle|^2.
\end{equation}

\section{Numerical Results}

We have implemented the teleportation protocol using the following temporal evolution schemes: (1) amplitude-unlimited optimization of $\varepsilon_0(t)$ and $\varepsilon_1(t)$; (2) amplitude-limited optimization of $\overline{\varepsilon}_0(t)$ and $\overline{\varepsilon}_1(t)$, i.e., ($0\leq\overline{\varepsilon}_k\leq1$, $k=0,1$); (3) single-control limited-optimization, i.e., $ \overline{\varepsilon}_1(t)=1-\overline{\varepsilon}_0(t)$ with ($0\leq\overline{\varepsilon}_0\leq1$; (4) QAOA approach and (5) linear adiabatic evolution (LAE), where $\varepsilon_0(t)=1-t/T$ and $\varepsilon_1(t)=t/T$. The LAE is also used to initialize the TBQCP method, see Eqs.~(\ref{oper}-\ref{fieldn}). 

\begin{figure}[th!]
\includegraphics{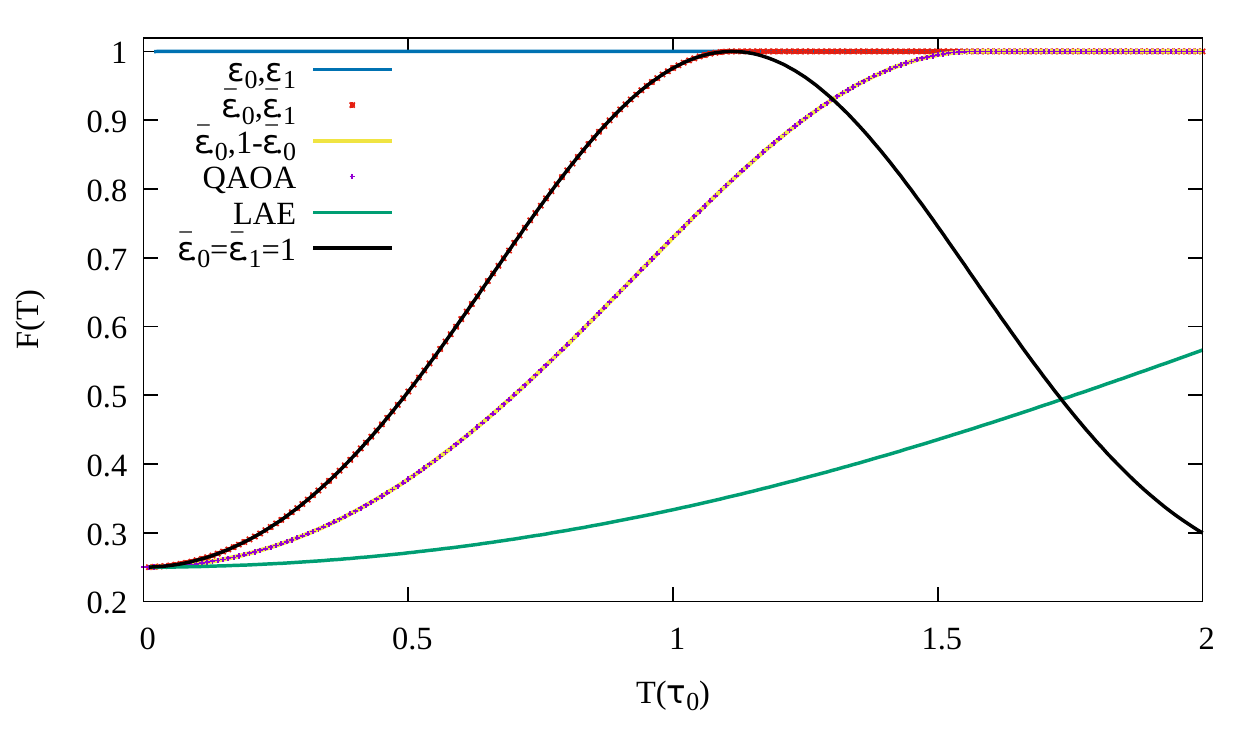}
    \caption{Fidelity $F(T)$ as a function of the final time of evolution, using the first (blue solid curve), the second (red dotted curve), the third (yellow solid curve), the fourth (magenta dotted curve), and the fifth (green solid curve) time evolution schemes. The black solid curve shows $F(T)$ as a function of the final time of evolution for the two-control functions set at their maximum value.}\label{fig1}
    \end{figure}

In Fig.~\ref{fig1}, we compare the fidelity as a function of final time $T$ considering different temporal evolution schemes. The first scheme achieves a fidelity equal to one in a very short final time because no limits are imposed on the two-control functions. This result shows that the amplitude-unlimited optimization is preferable if the goal is to implement the teleportation protocol in the fastest way. Considering the schemes where the controls are kept between the interval $[0,1]$, the one that achieves a higher fidelity is the amplitude-limited optimization, as can been noticed by the dotted curve in Fig.~\ref{fig1}. In this case, the fidelity reaches $1$ for final times above $T=1.11\tau_0$, where $\tau_0=\omega_0^{-1}$ is the time scale. The QAOA approach and the single-control limited-optimization essentially have the same performance and the fidelity reaches $1$ above $T=1.5\tau_0$. As expected, all schemes that employ an optimization procedure achieve a higher fidelity than the one achieved by the LAE. As demonstrated in section II, when the final time $T$ is not long enough to obtain the desired ground state $|\chi_0\rangle$, the optimal solution is accomplished by setting the two-control functions equal to the upper bound. The black solid curve in Fig.~\ref{fig1} shows $F(T)$ for different final times assuming that  $\overline{\varepsilon}_1(t)=\overline{\varepsilon}_0(t)=1$. One can notice that the fidelity for this case is equal to the amplitude-limited optimization up to the critical final time, which is $T_c=1.11\tau_0$ for the teleportation protocol. For $T>T_c$, the fidelity drops and reaches the minimum of 0.25 for $T=2T_c$. In this configuration, the fidelity can be well approximated by $F(T)\approx 0.25+0.75\sin^2{\left(\frac{\pi T}{2T_c}\right)}$, which reveals that the two-control functions set at the upper bound can perform a computational task with high-fidelity. The only requirement is to find the critical time $T_c$. Moreover, if the $T_c$ is too long to operate in a NISQ device, it be diminished by increasing the upper bound.

\begin{figure}[th!]
\includegraphics{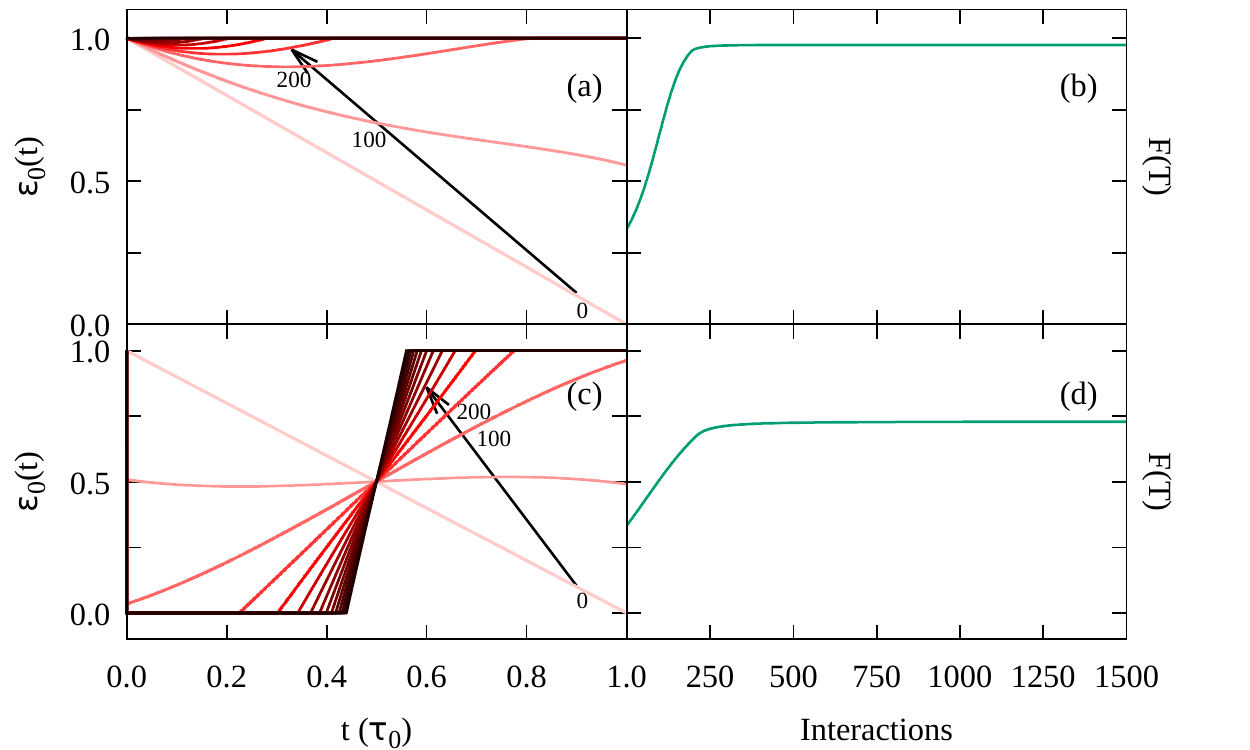}
    \caption{Panels (a) and (c) show the control $\varepsilon_0(t)$ for 0 to 1400 iterations of the TBQCP method, with step of 100 iterations, for the second and third schemes, respectively. The number of iterations increases in the direction indicated by the arrows and the color gradient. Thus, the darker the color, the higher the number of interaction in the TBQCP. Fidelity $F(T)$ as a function of the number of iterations in the TBQCP method, using the second and the third schemes are shown in panels (b) and (d).}\label{optimization_fig}
\end{figure}

Within the TBQCP method, we impose the bounds on the two-control functions by fixing the functions values in the maximum or minimum bound whenever the functions cross these limits. To better understand this idea, we plot $\overline{\varepsilon}_0(t)$ considering different iterations of the TBQCP in panel (a) of Fig.~\ref{optimization_fig} when the final time is $T=1.0\tau_0$. The initial trial function (interaction 0) is $\varepsilon_0(t)=1-t/T$. After 100 iterations, the function $\varepsilon_0(t)$ starts to bent up, yet does not cross any bound of the interval $[0,1]$. On the other hand, the TBQCP method finds functions that cross the upper bound after 200 iterations and we force the function to the maximum value within the region where the function cross the upper bound. The function $\overline{\varepsilon}_0(t)$ becomes constant and reaches the upper bound in the whole time evolution, as demonstrated in section II, after 1400 iterations of the TBQCP method. A similar behavior is found to $\overline{\varepsilon}_1(t)$ (not shown here). In panel (b) of Fig.~\ref{optimization_fig}, we show the fidelity $F(T)$ as a function of the TBQCP iterations, which converges to 0.9763 after 500 iterations. For all numerical calculation, we use $\eta=5\times10^{-3}$ of Eq.~(\ref{fieldn}). In panel (c) of Fig.~\ref{optimization_fig}, we plot the function $\overline{\varepsilon}_0(t)$ within the amplitude-limited optimization scheme as a function of the iterations of the TBQCP method, which cross the bounds after 300 iterations. After 1400 iterations, the function $\overline{\varepsilon}_0(t)$ resembles a step function, similarly to the functions obtained in the QAOA approach. Fidelity $F(T)$ as a function of the TBQCP iterations for the third scheme is shown in panel (d) of Fig.~\ref{optimization_fig}, which reaches convergence of 0.729 after 1000 iterations.

\begin{figure}[th!]
\includegraphics{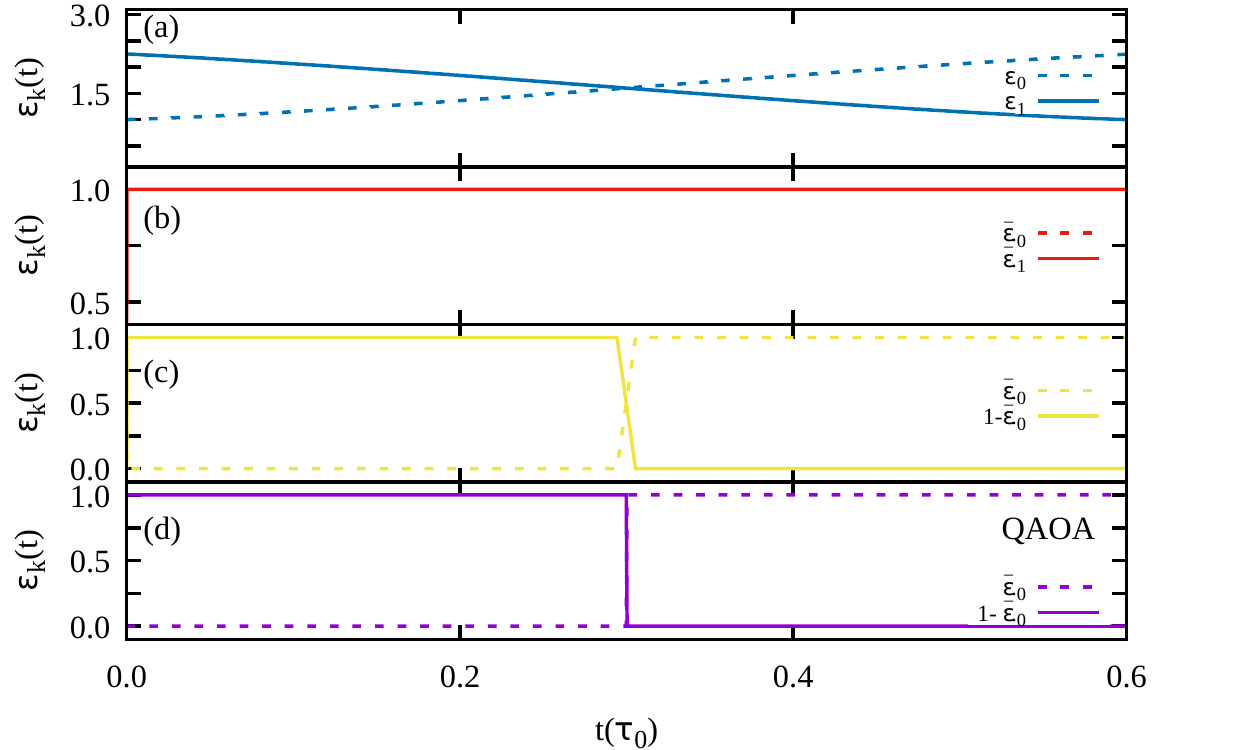}
    \caption{Optimized controls $\varepsilon_0(t)$ (solid lines) and $\varepsilon_1(t)$ (dashed lines) as a function of time for all considered schemes for the final time $T=0.6\tau_0$. The first to the fourth temporal evolution scheme is shown in panel (a) to panel (d),respectively.}\label{figT0.6}
\end{figure}

\begin{figure}[th!]
\includegraphics{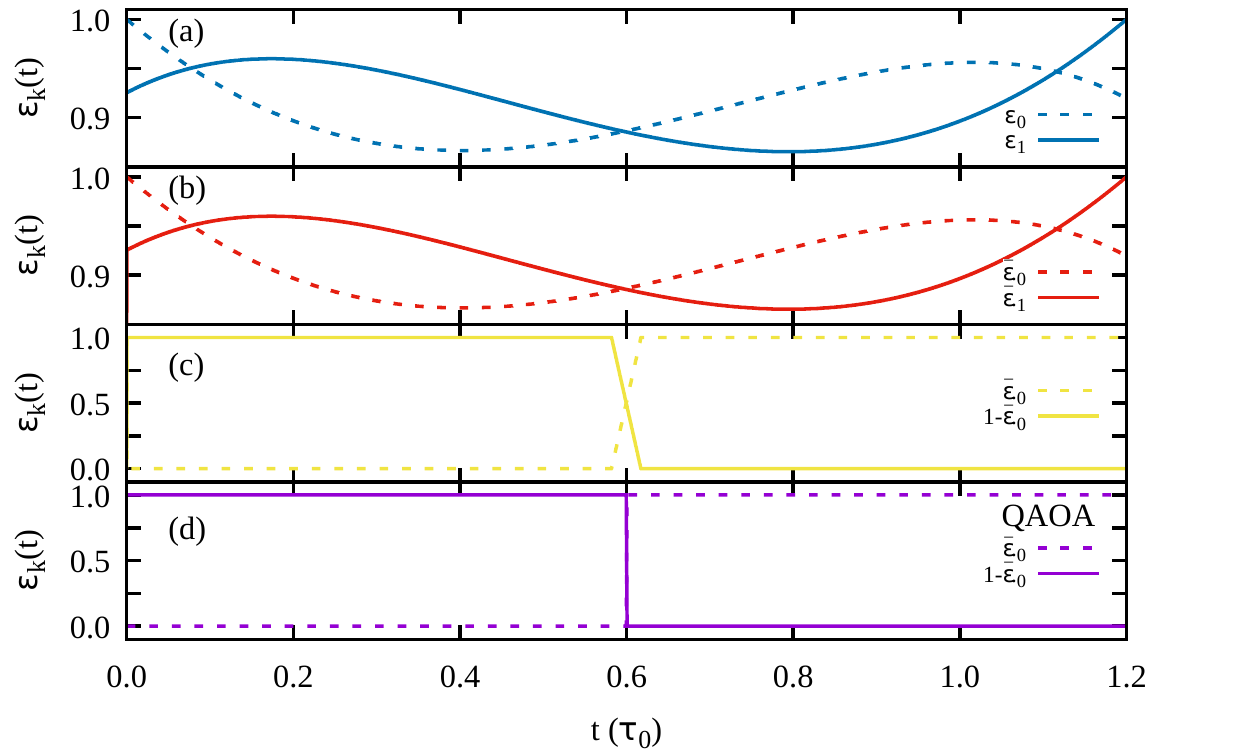}
    \caption{Optimized fields $\varepsilon_0(t)$ (solid lines) and $\varepsilon_1(t)$ (dashed lines) as a function of time for all considered schemes for the final time $T=1.2\tau_0$.The first to the fourth temporal evolution scheme is shown in panel (a) to panel (d),respectively. }\label{figT1.2}
 \end{figure}

\begin{figure}[th!]
\includegraphics{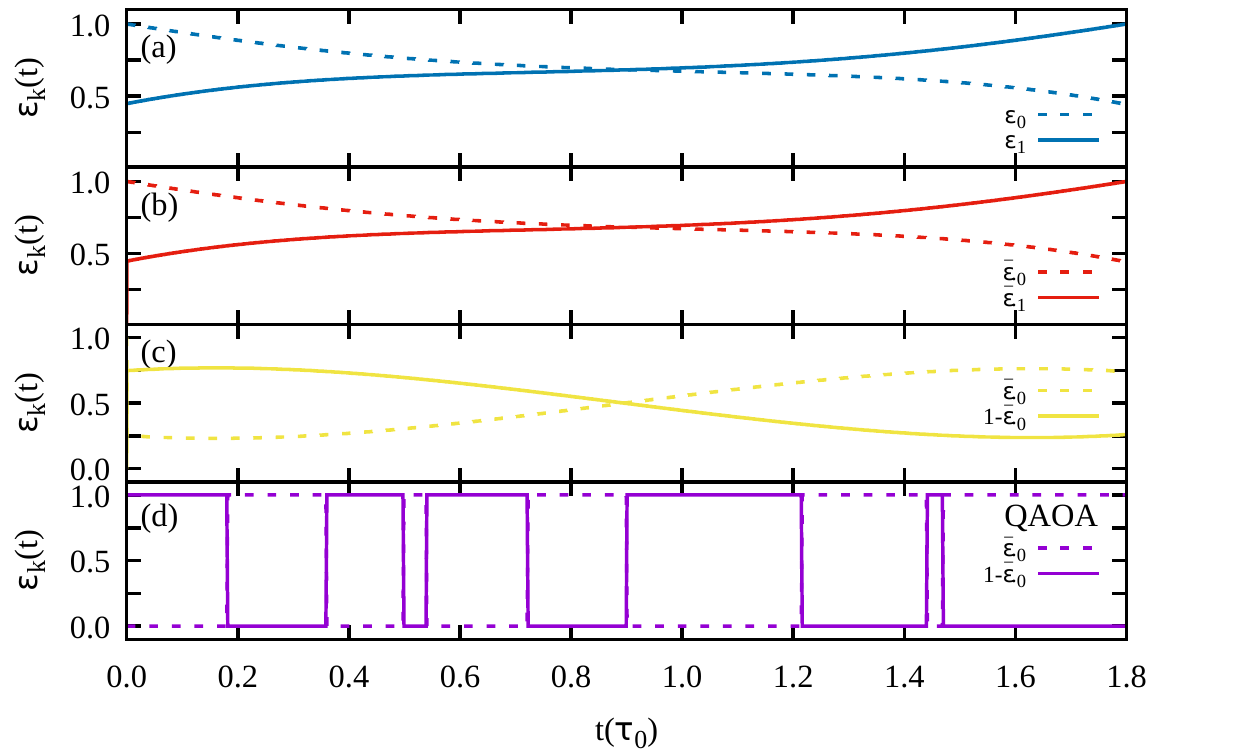}
    \caption{Optimized fields $\varepsilon_0(t)$ (solid lines) and $\varepsilon_1(t)$ (dashed lines) as a function of time for all considered schemes for the final time $T=1.8\tau_0$. The first to the fourth temporal evolution scheme is shown in panel (a) to panel (d),respectively.}\label{figT1.8}
 \end{figure}

In Figs. \ref{figT0.6} to \ref{figT1.8}, we show the converged optimal controls $\varepsilon_0(t)$ (solid lines) and $\varepsilon_1(t)$ (dashed lines) obtained for different final times. In each of these figures, panels (a) to (d) show the controls from the first to the fourth temporal evolution scheme, respectively. Observing panel (a) of Fig.~\ref{figT0.6}, $T=0.6\tau_0$, we note that the controls obtained in the amplitude-unlimited optimization exhibit an almost linear behavior and their amplitudes exceed one in the whole time interval. For this particular final time $T=0.6\tau_0$, only the first temporal evolution scheme is able to attain the maximum fidelity. In panel (b) of Fig.~\ref{figT0.6}, we observe that the solution for the amplitude-limited optimization corresponds to the two-controls constant in time with their amplitudes equal to the maximum limit, which is in accordance to the general results obtained in section II. The single-control limited-optimization and the QAOA result in similar solutions for the controls. This indicates that the QAOA approach with $p=1$ is very close to the optimal solution in the case of a single limited-amplitude control. It is also interesting to note that the schemes (1) to (3), initialized with the smooth linear ramp of LAE, converged to controls of opposite behavior compared to LAE, {\it e.g.}, $\varepsilon_0(0)>\varepsilon_1(0)$ in the LAE scheme, but $\varepsilon_1(0)>\varepsilon_0(0)$ for the optimization schemes when $T=0.6\tau_0$.

In Fig.~\ref{figT1.2}, we plot the resulting controls considering $T=1.2\tau_0$. Regarding the first and second schemes, we find that the two-controls functions are essentially the same because the optimal solutions do not cross the limits in the interval $[0,1]$. Again, the two-controls of the single-control limited-optimization are quite similar to the QAOA approach with $p=1$, where the difference is only the smoother transition between $0$ and $1$ for the third scheme. Figure \ref{figT1.8} shows the optimized controls for $T=1.8\tau_0$. We note that the single-control limited-optimization scheme has a different behavior as compared to the QAOA approach, while in the former there is only one jump at $T=0.9\tau_0$ the latter presents ten jumps. Although the controls obtained from different schemes have a different temporal evolution, the fidelity is always equals to $1$ for $T=1.8\tau_0$. This result is in agreement to the fact that the control landscape contains infinite optimal solutions \cite{PhysRevA.72.052337}. Another interesting information that we numerically verified is that the optimized controls functions can be very well approximated by a third-degree polynomial outside the regions where their values are set equal to the bounds. This fact evidences that these functions are simple and have the potential of being experimentally implemented.

\begin{figure}[th!]
\includegraphics{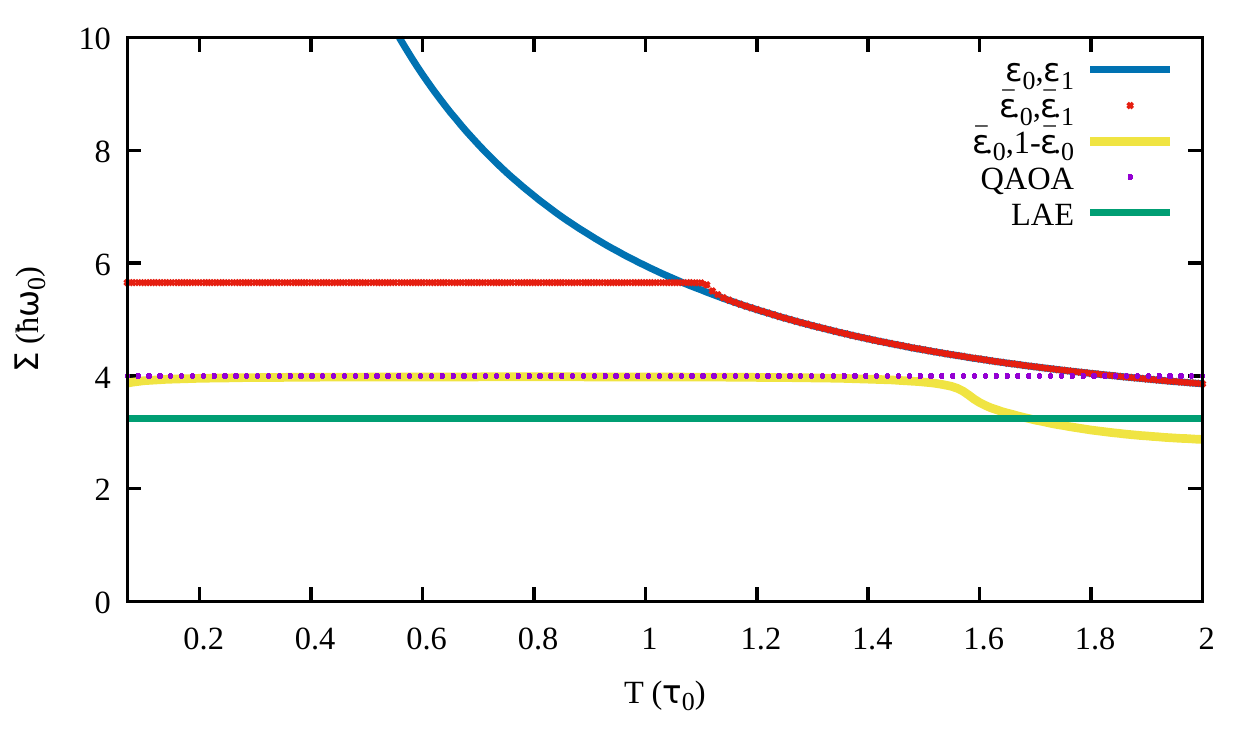}
    \caption{Energy cost as a function of the final time for the first (blue solid curve), the second (red dotted curve), the third (yellow solid curve), the fourth (magenta dotted curve), and the fifth (green solid curve) schemes of optimization.}\label{fig2}
 \end{figure}

We also perform the analysis of the energetic cost of implementing these different temporal evolution schemes. In Fig.~\ref{fig2}, we plot the energy cost as a function of the final time calculated according to~\cite{Alan_teleport}:
\begin{equation}
    \Sigma(T)=\frac{1}{T}\int_0^Tdt||H(t)||,
\end{equation}
where $||H(t)||=\sqrt{\mathrm{Tr}\left\{H(t)^2\right\}}$. We note that both the QAOA and the LAE schemes have constant energy cost for all $T$ shown in the Fig.~\ref{fig2}. As expected, the first temporal evolution scheme presents the higher energetic cost for T$<1.1\tau_0$. Above $T=1.1\tau_0$, the first and the second schemes have the same energetic cost because in this case the amplitude-unlimited optimization results in controls that are within the
interval $[0,1]$. Below $T=1.1\tau_0$, the second scheme has a constant energetic cost since the optimal solution is the controls constant equal to the upper bound. Above $T=1.5\tau_0$, the third scheme has a lower energetic cost than the QAOA approach and above $T=1.7\tau_0$ is even lower than the LAE, but with a considerably higher fidelity. Comparing the fidelity and the energy cost for the second and the third schemes in Figs. (\ref{fig1}) and (\ref{fig2}), we note that the values of $T$ for which the fidelity reaches one, correspond to the decreases in the energy cost of the respective schemes. As already shown, this is related to the change in the form of the corresponding controls.



\begin{figure}[th!]
\includegraphics{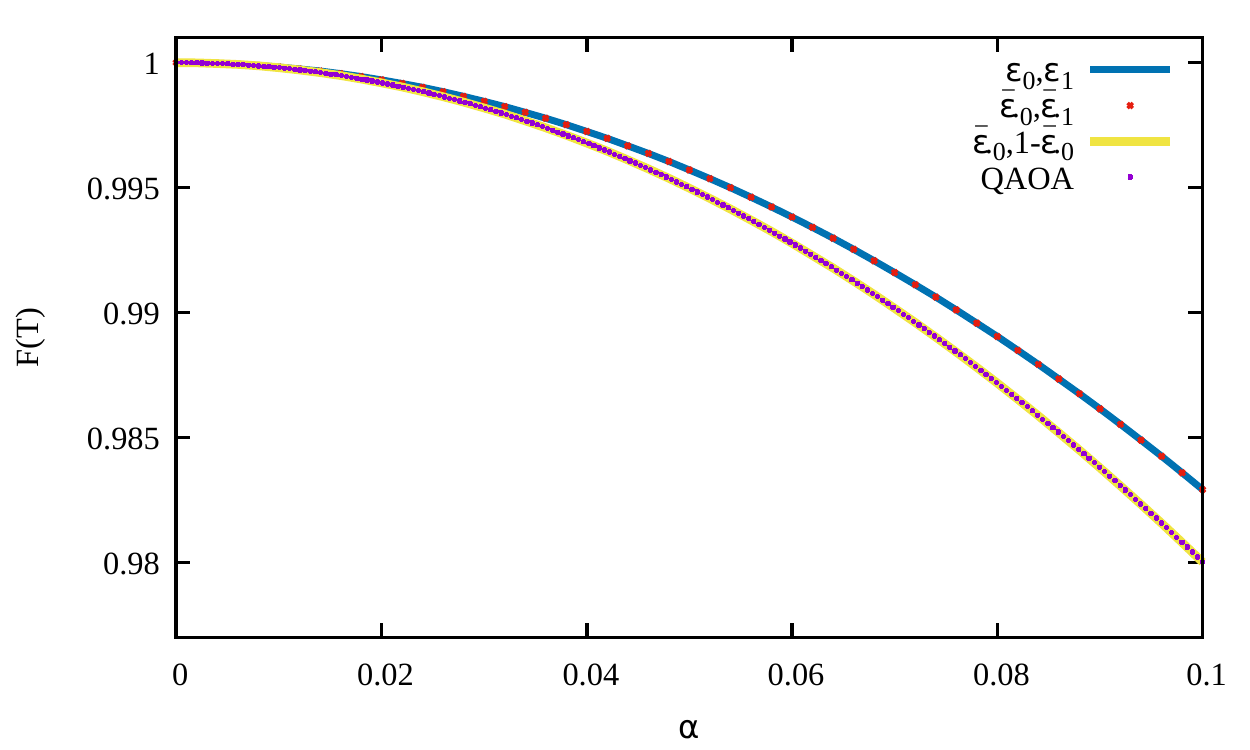}
    \caption{Largest systematic error as a function of the magnitude of a local magnetic field considering the first (blue solid curve), second (red dotted curve), third (yellow solid curve), and fourth (magenta dotted)  schemes of optimization.}\label{fig_error}
 \end{figure}

To probe the robustness of the optimization schemes against systematic errors, we have added an error Hamiltonian given by $H_e=\alpha\hbar\omega_0\sigma^j_k$ to equation~(\ref{Ht}), where $\alpha$ is proportional to the magnitude of the local magnetic field. We can find the most aggressive type of systematic error by testing all different combinations of $\sigma^j_k$ and keeping the one that affects more the fidelity. Figure~\ref{fig_error} shows the fidelity as a function of $\alpha$ considering $T=1.6\tau_0$, which is a final time where the fidelity reaches one for all optimization schemes, except LAE. The fidelity is evaluated using the optimized controls, but the time evolution is calculated including the error Hamiltonian $H_e$ in Eq.~(\ref{Ht}). Although, all schemes reach the maximum fidelity, we find that the first and second schemes are more robust against the systematic errors and a 10\% of error ($\alpha=0.1$) causes a reduction of 1.7\% in the fidelity. 


\section{Conclusions}

We have considered the quantum computing problem of reaching a target state by applying a combination of the problem-Hamiltonian and the driving-Hamiltonian, each one modulated by an independent control function. We have analytically demonstrated that when the final time is short such that the target state cannot be exactly prepared, the optimal solutions are those with both controls set to their maximum values during the whole evolution time. This type of solution is appealing due to its simplicity and the possibility of achieving the target state with high fidelity by increasing the final evolution time up to a critical value, where the fidelity approaches one. We also have shown that the use of two controls can generate higher yields than standard QA and QAOA schemes for the same evolution time. As an illustration, we have considered a Teleportation protocol fitted in the two-control formulation. The controls have been numerically obtained by the TBQCP method and also by the QAOA approach. We have confirmed the theoretical predictions finding that the two-controls scheme can perform the teleportation with a higher fidelity within the shortest time compared to the single control schemes. The price to pay is, of course the energy cost, which is higher for the two-control scheme. We also numerically confirmed that for small final times the optimal solution is the one with the two controls constants in time at their highest values, which we can term "double simultaneous bang" solution. Considering the schemes with a single control, we have found that QAOA and optimized control scheme have the same performance and similar bang-bang shape for small final times, but are quite distinct when the final time is sufficient to reach the target state.   

\section*{Acknowledgments}

We thank Alan Costa dos Santos for fruitful discussions and suggestions.
 EFL and LKC acknowledges support from S\~ao Paulo Research Foundation, FAPESP (grants 2014/23648-9 and 2019/09624-3) and from National Council for Scientific and Technological Development, CNPq (grants 423982/2018-4 and 311450/2019-9).

\pagebreak

    \pagebreak
    
  \pagebreak  
    
\pagebreak

\pagebreak

\pagebreak

\pagebreak

 \pagebreak
 
 \bibliography{quantumcomp}

\end{document}